\documentclass[10pt]{iopart} 

\usepackage{graphicx} 
\usepackage{soul} 
\usepackage{xcolor}

\usepackage[squaren]{SIunits} 

\newcommand{\neff} {n_{\mathrm{eff}}}
\newcommand{\nair} {n_{\mathrm{air}}} 
\newcommand{\lMC} {l_{\mathrm{MC}}} 
\newcommand{\wwg} {w_{\mathrm{wg}}} 
\newcommand{\DC} {\mathrm{DC}}

\newcommand{\etaldot} {$et$ $al$.} 

\begin{document}

\title{Design of free-space couplers for suspended triangular nano-beam waveguides}

\author{J.P. Hadden$^{1,2}$, Cobi Maynard$^{1}$, Daryl M. Beggs$^{1}$, Robert A. Taylor$^{3}$, Anthony J. Bennett$^{1,2}$}
\address{$^1$ School of Physics and Astronomy, Cardiff University, Queen’s Buildings, Cardiff CF24 3AA, United Kingdom}
\address{$^2$ School of Engineering, Cardiff University, Queen’s Buildings, Cardiff CF24 3AA, United Kingdom}
\address{$^3$ Clarendon Laboratory, Department of Physics, University of Oxford, Parks Road, Oxford OX1 3PU, United Kingdom}

\ead{haddenj@cardiff.ac.uk}

\begin{abstract}
Photonic waveguides with triangular cross section are being investigated for material systems such as diamond, glasses and gallium nitride, which lack easy options to create conventional rectangular nanophotonic waveguides. The design rules for optical elements in these triangular waveguides, such as couplers and gratings, are not well established. Here we present simulations of elements designed to couple light into, and out of, triangular waveguides from the vertical direction, which can be implemented with current angled-etch fabrication technology. The devices demonstrate coupling efficiencies approaching \unit{50}{\%} for light focused from a high numerical aperture objective. The implementation of such couplers will enable fast and efficient testing of closely spaced integrated circuit components. 
\end{abstract}

%
%
%
%
\ioptwocol

\section{Introduction}

Triangular waveguides have been used to study integrated photonics with colour centres hosted in diamond~\cite{Bayn2011, Burek2012}, silicon carbide\cite{Majety2021} and also with rare earth ions in glass~\cite{Ruskuc2022}. Angled etching of these materials creates high-index-contrast waveguides, and thus tight confinement of the optical field, albeit with geometry that is less well studied than conventional rectangular or ridge waveguides. Two methods have been used to create triangular waveguides: (1) Faraday cage-assisted dry-etching, which deflects the ions in a inductively coupled plasma etcher away from the surface normal, creating under-cut structures in a single step by etching from both sides simultaneously as shown by Burek \etaldot~\cite{Burek2012}, and (2) angled focused-ion-beam etching (FIB), where sequential etching from one side, then the other, fully releases the waveguide from the surrounding material as shown by Bayn \etaldot~\cite{Bayn2011}. Both methods are useful for materials systems without a selective anisotropic etch that that may otherwise be used to undercut. The latter can offer in-situ control of the waveguide position and shape, but the former is best suited to mass-manufacture of many devices in one step. 

\begin{figure*}[h]
    \centering
    \includegraphics[width=\textwidth]{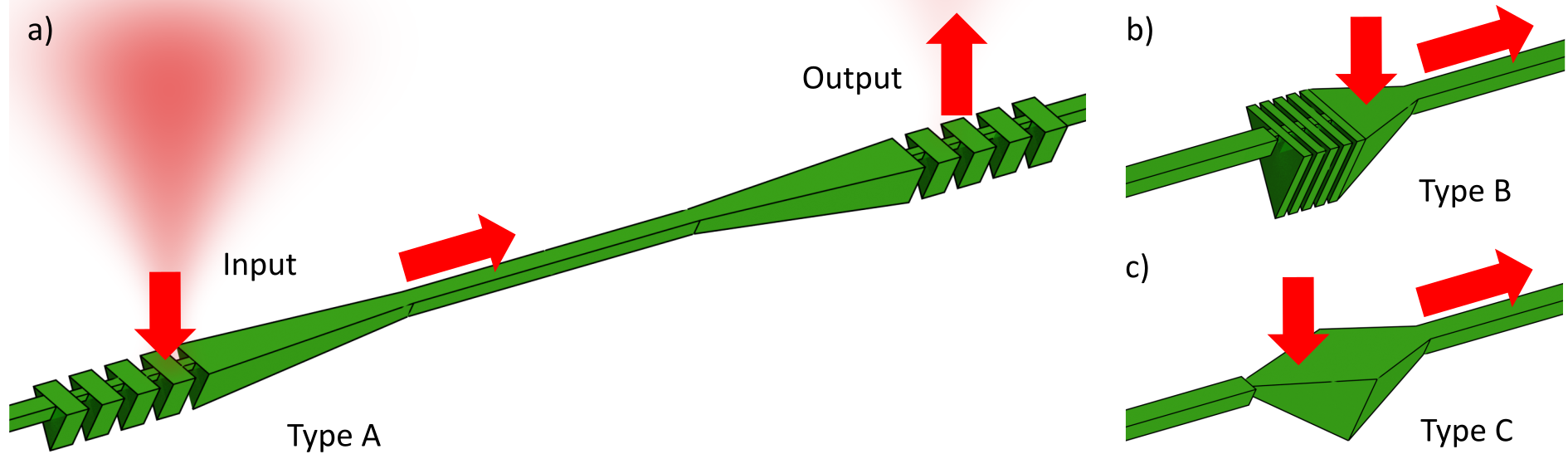}        
    \caption{a) Illustration of waveguide device with input and output couplers of Type A - Grating coupler with mode converter. b) Type B - Wedge mirror with Bragg reflector, and c) Type C - Wedge mirror with mode converter.}
    \label{fig:Figure_CouplerDesigns}
\end{figure*}

Recently, the fabrication of suspended triangular nano-beams in gallium nitride (GaN) has been demonstrated using angled Faraday cage-assisted etching~\cite{Gough2020}. GaN is a wide band-gap compound semiconductor underpinning high-power electronics and UV-visible light sources. It has also been shown to host single photon emitters with wavelengths spanning the visible to near infra red spectrum - which is a key requirement for optical quantum technologies~\cite{Berhane2017a,Zhou2017,Bishop2022}. New fabrication technology can open the way towards integrated optical devices based on this versatile electro-optic material, such as low-footprint nano-lasers~\cite{Niu2015}, non-linear optical light sources~\cite{Rahmani2018} and high speed optical switches~\cite{Cuniot-Ponsard2014,Ying2021} with potential applications as on-chip optical interconnects.

\begin{figure}[hb]
    \centering
    \includegraphics[width=0.5\textwidth]{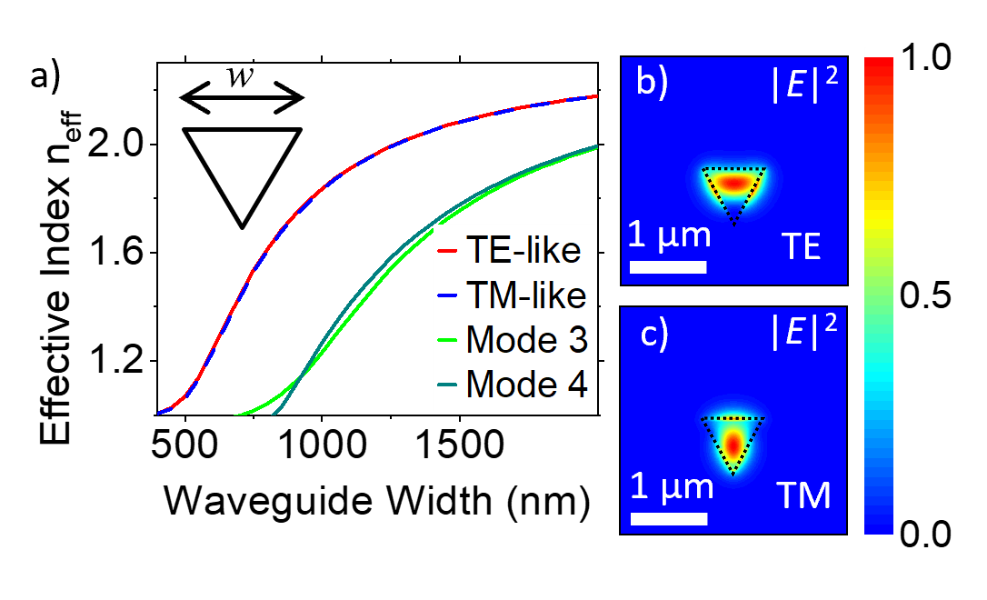}    
    \caption{a) Refractive index of triangular waveguides as a function of top waveguide width. b) Electric field intensity of TE and c) TM -like modes for a waveguide with top width of \unit{765}{\nano\meter}} 
    \label{fig:Figure_RefractiveIndex}
\end{figure}

Light may be coupled into and out of waveguides using grating couplers based on periodic modulation of the effective refractive index.  Such vertical couplers have important applications in the testing and development of photonic integrated circuit components, allowing for fast and efficient testing of closely spaced components. In ridge waveguides, such as silicon on insulator, this is commonly achieved by etching periodic trenches into the waveguide. The effective refractive index is dependent on the etch depth and length of the etched sections, reducing the effective refractive index $\neff$ of the waveguide mode in these regions~\cite{Marchetti2017}. For triangular waveguides, one may modify the refractive index by modulating the top waveguide width. Figure \ref{fig:Figure_RefractiveIndex} a) shows the effective refractive indices of the first four modes in an equilateral triangular waveguide as a function of the waveguide top width width $w$, at \unit{1550}{\nano\meter} wavelength, calculated using Ansys Lumerical Mode~\cite{Lumerical}. The first two modes may be categorised as TE-like and TM-like, while the subsequent modes appear to show rotation symmetry of order three. The nominal target waveguide size is chosen to be \unit{765}{\nano\meter}, the largest size that prevents modes 3 and 4 being guided ~\cite{Maynard2022}. Field intensity plots of the TE and TM modes for a waveguide with top width of \unit{765}{\nano\meter} are shown in Figure \ref{fig:Figure_RefractiveIndex} b) and c).

Although there has already been considerable effort developing optimised fibre to chip couplers in conventional rectangular waveguide geometries~\cite{Cheng2020}, there has been much less on the on the problem of coupling light from microscope objectives into nanophotonic waveguides in or in triangular waveguide geometries. In this work we present designs for free-space couplers based on the triangular nano-beam geometry which couple light focused from above to a Gaussian spot by a high NA microscope objective. We present three classes of designs as illustrated in figure \ref{fig:Figure_CouplerDesigns}: Type A - grating coupler with mode converter, Type B - wedge mirror with Bragg reflector, and Type C - wedge mirror with mode converter, with performance of the designs is evaluated and optimised using Ansys Lumerical FDTD~\cite{Lumerical}. We focus on gallium nitride as the host material. Although the refractive indices may be different, the design concepts and fabrication processes will be transferable to other material systems such as glass, diamond, silicon, silica and other compound semiconductors.

\section{Type A coupler design - Grating coupler with mode converter}
\begin{figure*}[t]
    \centering
    \includegraphics[width=\textwidth]{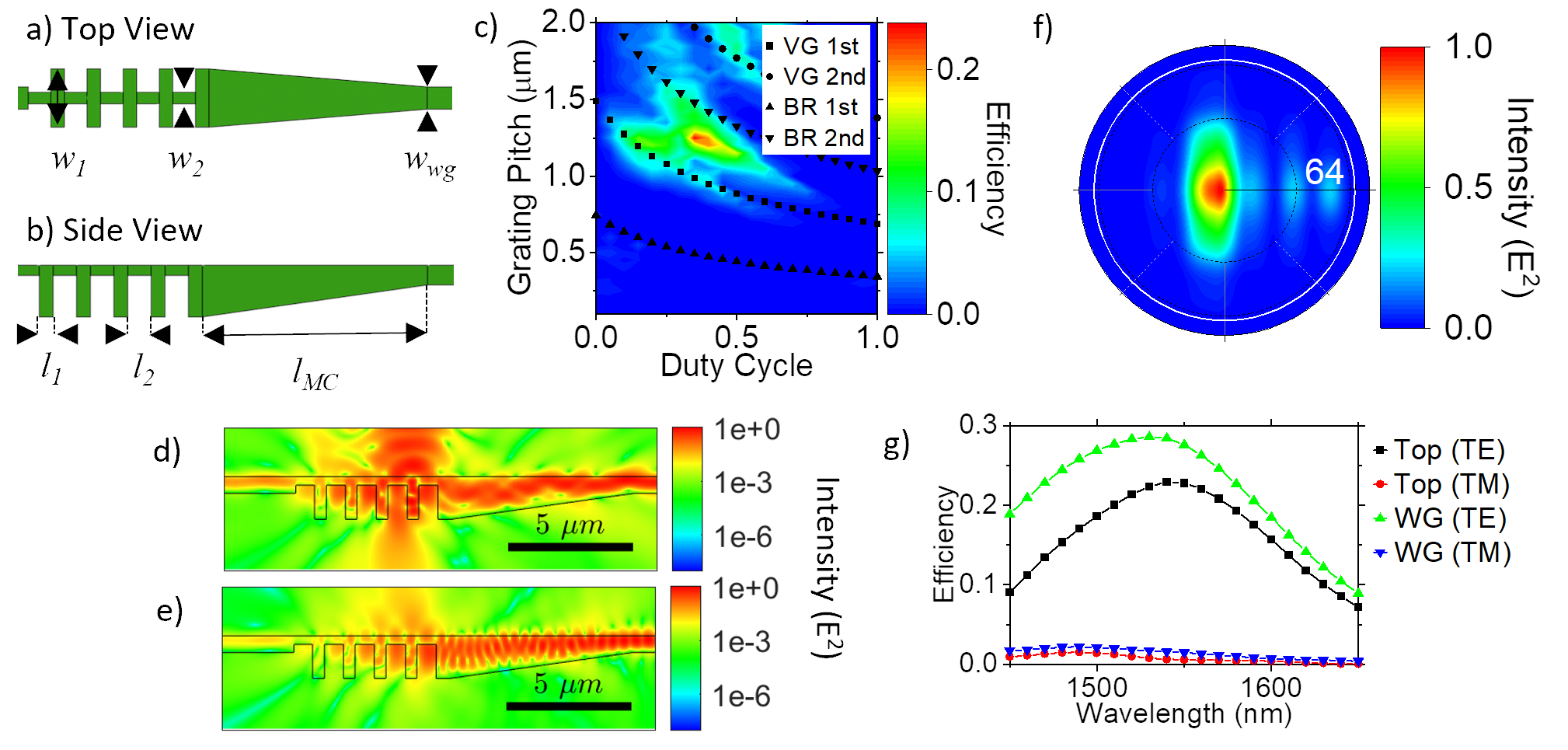}
    \caption{\textbf{Type A coupler design} a) Top and b) side view with main design parameters marked: grating coupler widths ($w_1$ and $w_2$) and lengths ($l_1$ and $l_2$), mode converter length ($\lMC$) and waveguide width ($\wwg$). c) Calculated coupling efficiency for a Gaussian source focused from above into the TE-like waveguide mode as function of duty cycle ($l_1/(l_1+l_2)$) and grating pitch ($l_1+l_2$) for a grating coupler with $w_1 = \unit{2000}{\nano\meter}$ and $w_2 = \unit{400}{\nano\meter}$ The pitch and duty cycle ($\DC$) which satisfy the 1st and 2nd order vertical grating (VG) and Bragg reflector (BR) phase matching conditions are marked. d) Side view cross section field intensity distribution for coupler with light input from the top and e) through the waveguide with TE-like polarisation. f) Far field intensity distribution with TE waveguide input source. 
    g) Simulated coupling efficiency as a function of wavelength for light input from the top (labelled "Top") and through the waveguide (labelled "WG") in TE and TM-like polarisations.}
    \label{fig:Figure_TypeA}
\end{figure*}

The top and side profiles of a Type A vertical grating coupler designed to couple light focused from above using a 0.9 NA objective lens are shown in Figure \ref{fig:Figure_TypeA} a) and b). The grating region consists of repeating sections of thicker (width $w_1 = \unit{2000}{\nano\meter}$, length $l_1$) and thinner (width $w_2 = \unit{400}{\nano\meter}$, length $l_2$) sections, with a mode converter (length $\lMC=\unit{7.5}{\micro\meter}$) allowing adiabatic transition between the thicker grating region, and the $\wwg=\unit{765}{\nano\meter}$ waveguide output. The phase matching condition for a grating coupler is $\neff - \nair \sin(\theta ) = m ~ \lambda / a$, where $\neff$ is the effective index of the optical mode in the grating region, $\nair$ is the refractive index of air, $\theta \approx \unit{0}{\degree}$ is the target grating diffraction angle, $m$ is the grating order number, $\lambda$ is the wavelength of the light, and $a = l_1 + l_2$ is the grating pitch. The effective refractive index of the optical mode in the grating region may be estimated using $\neff = \sqrt{\DC~{n_1}^2 + ((1-\DC)~ {n_2}^2 }$ for TE Modes~\cite{Zhang2015c}, where $n_1$ and $n_2$ are the effective refractive indices for waveguide modes in the thicker ($w_1$) and thinner ($w_2$) sections respectively, and $\DC = l_1/(l_1 + l_2)$ is the duty cycle of the thicker ($w_1$) section. The coupling efficiency for a Gaussian source focused from above into the TE-like waveguide mode is calculated as a function of the grating duty cycle and grating pitch, as shown in Figure \ref{fig:Figure_TypeA} c). The efficiency is defined as the overlap of the normalised field intensity with the nano-beam's TE waveguide mode. The pitch and duty cycle which satisfy the 1st and 2nd order vertical grating and Bragg reflector phase matching conditions are marked as a dotted line in c). 

The most efficient design is achieved with duty cycle $\DC = 0.35$, and grating pitch $a = $ \unit{1.25}{\micro\meter}, leading to a coupling efficiency of \unit{24}{\%}. This is modest compared to grating couplers optimised for single mode optical fibre to chip coupling in silicon on silica platforms, where theoretical coupling efficiencies can exceed  \unit{90}{\%}~\cite{Bozzola2015}. However, such optical fibre grating couplers typically have dimensions comparable to the single mode fibres - orders of magnitude larger than the $\sim \unit{1}{\micro\meter^2}$ focussed spot considered here, and require long tapering regions to convert from the grating coupler to the waveguides.  Compact grating couplers designed for microscope collection from silicon nitride waveguides show more comparable theoretical coupling efficiencies of up to \unit{40}{\%} (or \unit{60}{\%} when including a reflective mirror underneath) \cite{Zhu2017}.  

Interestingly, this design does does not appear to satisfy the grating coupler phase matching condition. Designs which do are limited to around $\unit{15}{\%}$. We believe the reason that this design outperforms grating designs that satisfy the phase matching condition is that the size of focused the Gaussian spot is only large enough to cover one period of the grating, and thus, the device is exhibiting a Mie resonance which is enhancing the coupling - as recently shown in silicon nano-wire super-lattices~\cite{Kim2018}. Grating couplers which demonstrate out-coupling from a single period have also been demonstrated in gallium arsenide for out coupling quantum dot emission ~\cite{Faraon2008}. 

The grating coupler design was further optimised using a particle swarm algorithm, varying the grating pitch, duty cycle, and the offset of the focused source TE polarised source. The resulting solution was then evaluated using four different input scenarios: Top (TE) focused Gaussian input with TE-like polarisation - where the electric field is perpendicular to the waveguide, Top (TM) focused Gaussian input with TM-like polarisation - where the electric field parallel is to the waveguide, WG (TE) input through the waveguide TE-like mode, WG (TM) input through the waveguide TM-like mode. This allows us to test how the optimised coupler performs to couple light from free-space into the waveguide, and conversely from the waveguide into free-space in both TE and TE-like polarisations. Figure \ref{fig:Figure_TypeA} d) shows the field intensity of a cross section of the optimised grating coupler when illuminated from above with a TE source, demonstrating how the field is coupled from a focused Gaussian source above the grating region, to the waveguide output on the right hand side, while Figure \ref{fig:Figure_TypeA} d) shows the shows the field intensity when the waveguide is excited through the waveguide mode, while Figure \ref{fig:Figure_TypeA} f) shows a far field projection of the output of the coupler in the vertical direction. We can estimate the collection efficiency of this output into a microscope objective by integrating the field intensity within the angular cone defined by a 0.9 NA collection objective ($\theta = \unit{64}{\degree}$).  The performance of the coupler for both TE and TM-like modes from the four different input scenarios over a wavelength range of \unit{200}{\nano\meter} around the \unit{1550}{\nano\meter} design wavelength is shown in Figure \ref{fig:Figure_TypeA} g). It can be seen that the device performs well for TE-like modes with peak efficiencies of \unit{23}{\%} for top illumination, \unit{28}{\%} for waveguide illumination with a bandwidth of around \unit{200}{\nano\meter}, however for TM polarised light efficiencies peak at less than \unit{2}{\%}.

\section{Type B coupler design: Wedge mirror with Bragg reflector}
\begin{figure*}
    \centering
    \includegraphics[width=\textwidth]{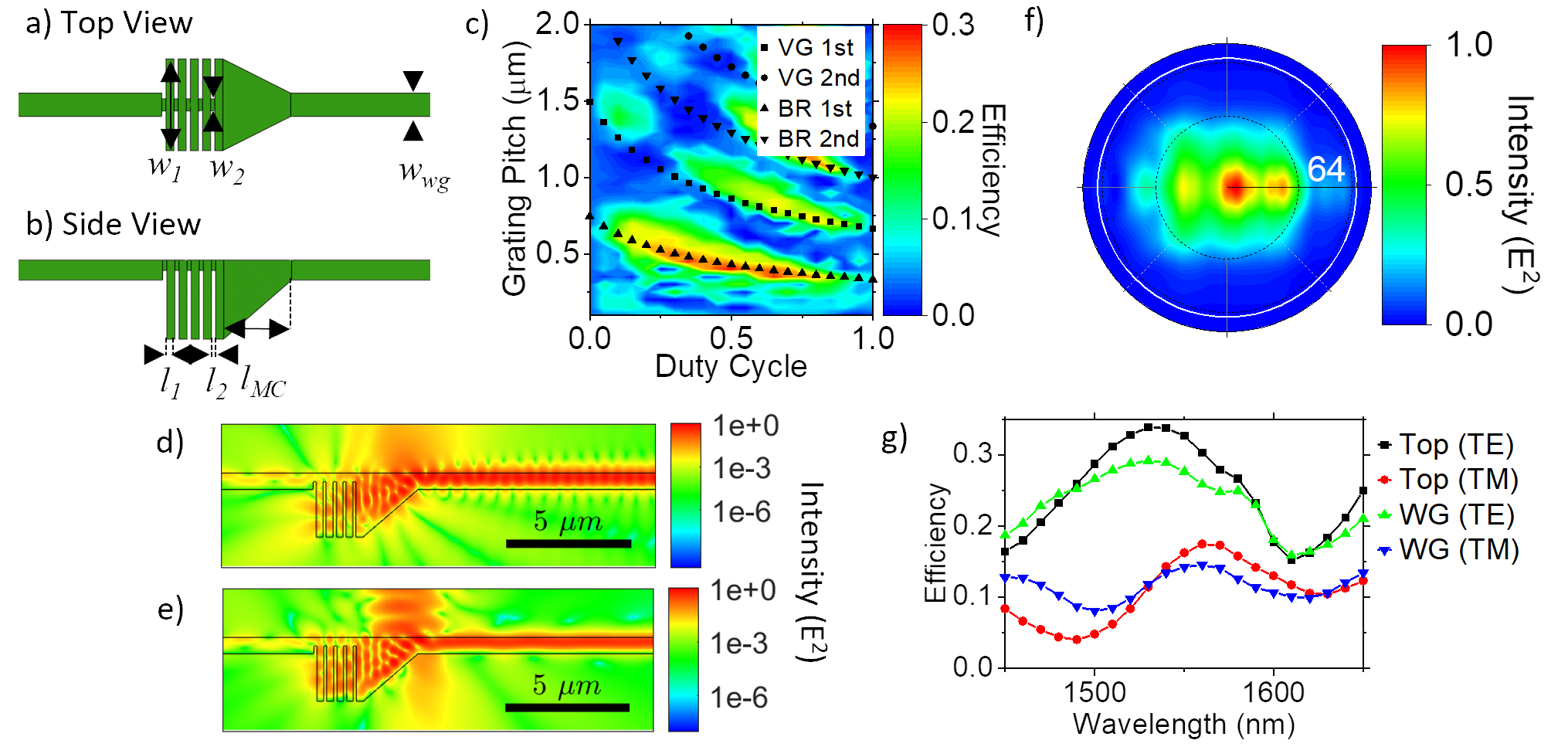}    
    \caption{\textbf{Type B coupler design} a) Top and b) side view with main design parameters marked: grating coupler widths ($w_1$ and $w_2$) and lengths ($l_1$ and $l_2$), mode converter length ($\lMC$) and waveguide width ($\wwg$). c) Calculated efficiency of coupling a Gaussian source focused from above into the TE-like waveguide mode as function of duty cycle ($l_1/(l_1+l_2)$) and grating pitch ($l_1+l_2$) for a Bragg reflector with $w_1 = \unit{3000}{\nano\meter}$ and $w_2 = \unit{400}{\nano\meter}$ The pitch and duty cycle which satisfy the vertical grating and Bragg reflector phase matching conditions are marked. d) Side view cross section field intensity distribution for coupler with focused Gaussian TE input and e) with TE waveguide input sources. f) Far field intensity distribution with TE waveguide input.
    g) Simulated coupling efficiency as a function of wavelength for light input from the top (Top) and through the waveguide (WG) in TE and TM-like polarisations.
    }    
    \label{fig:Figure_TypeB}
\end{figure*}

During optimisation of the type A design it was discovered that for certain lengths of $\lMC$ the mode converter section of the waveguide actually performs as an angled mirror, allowing light to be reflected through total internal reflection from the vertical direction to the horizontal. Rectangular total internal reflection mirror couplers have previously been demonstrated in diamond through focused ion beam milling~\cite{Castelletto2011}. However, the triangular geometry investigated here presents a tapered 'wedge mirror' shape. When combined with a Bragg reflector, it is possible to design a device which reflects and directs light focused from above into the the waveguide mode. 

Figure \ref{fig:Figure_TypeB} a) and b) show the top and side profiles of a Type B device combining a wedge mirror and a Bragg reflector. Bragg reflectors based on modulating the waveguide width can be implemented in a similar way to the grating couplers, however in this case the phase matching condition becomes $\neff = 2(m + 1)~ \lambda/2a$. The grating region consists of repeating sections of thicker (width $w_1 = \unit{3000}{\nano\meter}$, length $l_1$) and thinner (width $w_2 = \unit{400}{\nano\meter}$, length $l_2$) sections, with a wedge mirror (length $\lMC=\unit{2.25}{\micro\meter}$) allowing total internal reflection of the focused beam from above, and directing the reflected light into the $\wwg=\unit{765}{\nano\meter}$ waveguide output.

Figure \ref{fig:Figure_TypeB} c) shows the coupling efficiency of a Gaussian source focused from above into the TE-like waveguide mode as a function of the grating duty cycle, and grating pitch. The pitch and duty cycle which satisfy the vertical grating and Bragg reflector phase matching conditions are marked as a dotted line in c). It can be seen that the most efficient designs run along the Bragg reflector first order line as expected, with the best design parameters $DC = 0.5$ and grating pitch $a = \unit{0.4}{\micro\meter}$ showing efficiency of \unit{30}{\%}. It is notable that this is higher than the best Type A grating coupler design even though it combines two reflective elements.

Following a similar procedure as for the previous design, we optimise the coupling efficiency by varying the grating pitch, duty cycle, wedge mirror length and the offset of the device with respect to the focused source TE polarised source. We then evaluate the performance of the optimised coupler device as before for TE and TM polarised light for in-coupling a focused Gaussian beam from above into the waveguide mode (Top (TE) and Top (TM)), and for out-coupling the waveguide modes into a 0.9 NA collection cone representing a microscope objective (WG (TE) and WG (TM)). Figure \ref{fig:Figure_TypeB} d), and \ref{fig:Figure_TypeB} e), show the  field intensity cross sections for the Top (TE) and WG (TE) scenarios respectively, while \ref{fig:Figure_TypeB} f) shows the far field projection of the field intensity for the out coupled mode. Figure \ref{fig:Figure_TypeB} g) shows the performance of the device over a \unit{200}{\nano\meter} wavelength range around the design wavelength. 

The peak efficiency of this Type B design is \unit{35}{\%} for top illumination, and \unit{29}{\%} for waveguide illumination for TE polarised light, which is better than the Type A design. It also performs better than the Type A design for TM polarised light, with peak efficiencies of \unit{17}{\%} and \unit{14}{\%} respectively for Top (TM) and WG (TM) input scenarios, and spatially it is much more compact - with a total length of \unit{4.2}{\micro\meter} as compared to \unit{13.8}{\micro\meter} for design A, however the efficiency profile over the wavelength range, and the output mode are both more complex, indicating several output modes, which is less desirable for some applications - for example where the output needs to be coupled into a single mode fibre.


\section{Type C coupler design: Wedge mirror with mode converter}
\begin{figure*}
    \centering
    \includegraphics[width=\textwidth]{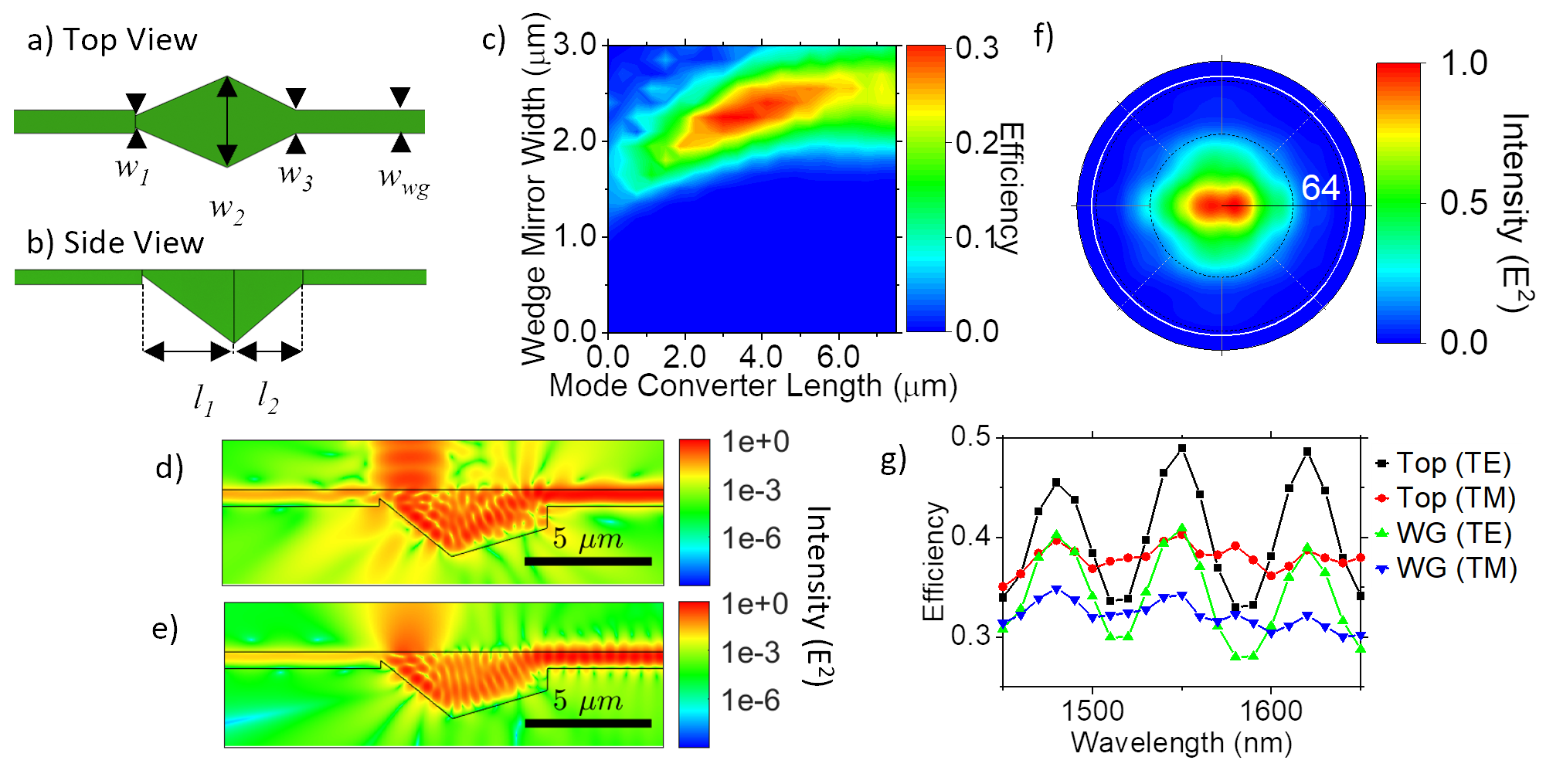}
    \caption{\textbf{Type C coupler design} a) Top and b) side view with main design parameters marked: wedge mirror and mode converter widths ($w_1$, $w_2$, $w_3$) and lengths ($l_1$, $l_2$) and waveguide width ($\wwg$). c) Calculated efficiency of coupling a Gaussian source focused from above into the TE-like waveguide mode as function of mode converter length and wedge mirror width $w_1 = \unit{2000}{\nano\meter}$ and $w_2 = \unit{2000}{\nano\meter}$ d) Side view cross section field intensity distribution for coupler with focused Gaussian TE input and e) with TE waveguide input sources. f) Far field intensity distribution with TE waveguide input source. 
    g) Coupling efficiency of coupler as a function of wavelength with focused Gaussian, and waveguide TE and TM like input sources.}
    \label{fig:Figure_TypeC}
\end{figure*}

The third design we consider, Type C, consists of a wedge mirror combined with a mode converter. This design uses only wedged mirrors and mode converters, with geometry controlled by their relative widths. Figure \ref{fig:Figure_TypeC} a) and b) show the top and side profiles of the Type C device, combining a wedge mirror with widths ($w_1 = \unit{400}{\nano\meter}$ and $w_2 = \unit{3000}{\nano\meter}$) and length $l_1$ and a mode converter with widths ( $w_2 = \unit{3000}{\nano\meter}$ and  $w_3 = \unit{765}{\nano\meter}$) and length $l_2$. Figure \ref{fig:Figure_TypeC} c) shows the coupling efficiency of a Gaussian source focused from above as the mode converter length ($l_2$) and wedge mirror/mode converter thicker width ($w_2$) length are varied. The highest efficiency design is obtained with $l_2 = \unit{3.75}{\micro\meter}$ and  $w_2 = \unit{2.25}{\micro\meter}$.

Subsequently, we optimise the design, allowing the wedge mirror output width ($w_2$), and length ($l_1$), and mode converter output width ($w_2$) and length ($l_2$), along with the offset of the device with respect to the focused Gaussian TE polarised source to be varied. This yields a device with a peak coupling efficiency of \unit{49}{\%} for a focused Gaussian TE beam. Interestingly, the highest efficiency design presents a step discontinuity between the mode converter and the output waveguide. This device shows good performance both for in-coupling (Figure \ref{fig:Figure_TypeC} d) and out coupling (Figure \ref{fig:Figure_TypeC} e), with a concentrated far field projection field intensity distribution (Figure \ref{fig:Figure_TypeC} f). It can also be seen in Figure  \ref{fig:Figure_TypeC} g) that although there are some spectral resonances visible, the device works relatively well for both TE and TM polarised light, in both in-coupling and out-coupling configurations, with coupling efficiencies greater than \unit{26}{\%} for all configurations across the range. 

\section{Conclusion}

In this work, we present the design of free space couplers for suspended triangular nano-beam waveguides. We investigate three different designs which demonstrate grating coupler, Bragg reflector, and total internal reflection elements, using the suspended triangular waveguide geometry. The optimised device designs demonstrate simulated coupling efficiencies approaching \unit{50}{\%} for light focussed from a high numerical aperture objective. The development of such couplers will enable fast and efficient testing of closely spaced integrated circuit components.

\section*{Acknowledgements}

This work was performed using the computational facilities of the Advanced Research Computing @ Cardiff (ARCCA) Division, Cardiff University. We acknowledge financial support provided by EPSRC via Grant No. EP/T017813/1 and EP/P006973/1.  

\section*{Data Availability}
The data that support the findings of this study are openly available
in the Cardiff University Research Portal at http://doi.org/10.17035/d.2022.0214949311.


\bibliographystyle{ieeetr} 
\bibliography{references}

\end{document}